\documentclass[aps,prl,epsfigure,showpacs,twocolumn]{revtex4}
\usepackage{dcolumn}
\usepackage{bm}
\usepackage{graphicx}
\usepackage{amsmath}
\usepackage{latexsym}
\usepackage{amsfonts}
\usepackage{amssymb}
\usepackage{array}
\usepackage{epsfig}
\usepackage{times}

\newcommand{\ket}[1]{\left\vert#1\right\rangle}

\newcommand{\miniproj}[2]{\vert#1\rangle\langle#2\vert}

\newcommand{\minivalmed}[1]{\langle#1\rangle}

\newcommand{\nbar}{\overline{n}}

\begin{document}

\title{Hamiltonian tomography in an access-limited setting without state initialization}

\author{C. Di Franco, M. Paternostro, and M. S. Kim}

\affiliation{School of Mathematics and Physics, Queen's University, Belfast BT7 1NN, United Kingdom}

\date{\today}

\begin{abstract}
We propose a scheme for the determination of the coupling parameters in a chain of interacting spins. This requires only time-resolved measurements over a single particle, simple data post-processing and no state initialization or prior knowledge of the state of the chain. The protocol fits well into the context of quantum-dynamics characterization and is efficient even when the spin-chain is affected by general dissipative and dephasing channels. We illustrate the performance of the scheme by analyzing explicit examples and discuss possible extensions.
\end{abstract}


\maketitle

It is often the case in the dynamics of interacting many-body systems that a specific and desired effect is achieved by means of an appropriately designed set of coupling strengths. Such effects are frequently very sensitive to even small deviations from the required pattern of interaction parameters, which may result in a dramatic deterioration of the performances. Moreover, in many cases the couplings are assumed to have been pre-engineered by a third party and it would be highly desirable, from a practical viewpoint, to test if we have been provided with the proper set of parameters before running a protocol. In other words, it would be important to have a {\it diagnostic} and {\it non-invasive} routine which allows one to infer the pattern of interaction strengths in a quantum many-body system with a high degree of accuracy. In essence, we would require the performance of a ``Hamiltonian tomography scheme". This can be seen as a variation of quantum process tomography~\cite{nielsenchuang} which, together with state~\cite{QST} and detector tomography~\cite{QDT}, allows for the complete characterization of quantum dynamics. All of them have found experimental verification. Our scheme greatly reduces the resources necessary to estimate the coupling parameters of the interaction model.

Reconstructing the form of a given but undisclosed one- and two-spin Hamiltonian has raised the interest of the physics community~\cite{schirmer} and, very recently, an intriguing proposal has been put forward for the $N$-particle case~\cite{burgarth}. Within the frameworks of these investigations, protocols able to find the coefficients characterizing the interaction Hamiltonian have been developed. However, {\it the initialization of the state of the system is required in each of them}. In addition, a complete set of relevant eigenvalues of the interaction model has to be known {\it a priori} or should be determined in an adept way, which may require the enforcement of strong conservation laws on the class of Hamiltonians that can be tested~\cite{burgarth}. These requirements are in general difficult to be met or unnecessarily limiting. In this Letter, we use an approach based on the ``information flux'' (IF)~\cite{informationflux} to investigate Hamiltonian tomography performed with minimal access to the many-body system and without the necessity for initial preparation~\cite{roastedchicken}. Moreover, we stress another remarkable advantage in the protocol we suggest: {\it the Hamiltonian to study does not need to commute with the total spin-excitation number}; {\it i.e.}, we do not require that the total number of excitations in the system is preserved. In clear contrast even with classical schemes for {\it inverse problems in vibration}~\cite{gladwell}, our method does not rely on the prior knowledge of a set of eigenvalues of the Hamiltonian. Information about the coupling coefficients is found via time-resolved single-spin measurements without state initialization of the system, which is distinctive and original, compared to what has previously been done~\cite{burgarth,gladwell}.

The difference with respect to quantum process tomography is also evident: there, {\it the initialization of the whole system} in a set of relevant states and the performance of state tomography after the action of the process are required. From the reconstructed output density matrices, one can then infer the completely positive map corresponding to the process itself. On the other hand, in our scheme, we just need to measure a single element of a multipartite register at various times; no condition on the state of the rest of the system is imposed. The time evolution of the expectation value of operators acting on that single spin can be extracted from the acquired data and the complete set of coupling coefficients of the Hamiltonian can be reconstructed from it.

To fix the ideas and clearly elucidate the main features of our study, we start from a simple excitation-preserving class of interaction models. We consider a linear chain of $N$ spin-$1/2$ particles, mutually coupled via the nearest-neighbor anti-ferromagnetic $XX$ Hamiltonian~\cite{antiferromagnetic}
\begin{equation}
\label{modelloXX}
\hat{{\cal H}}_1=\sum^{N-1}_{i=1}J_{i}(\hat{X}_{i}\hat{X}_{i+1}+\hat{Y}_{i}\hat{Y}_{i+1}).
\end{equation}
Here, $J_i>0$ is the interaction strength between spins $i$ and $i+1$ while $\hat{X}_i$, $\hat{Y}_i$ and $\hat{Z}_i$ denote the $x$, $y$ and $z$-Pauli matrix of spin $i$, respectively. {While it is important to remark, at this stage, that our method can be adapted to a larger class of Hamiltonians, as discussed later on, we clarify here that the choice of $\hat{{\cal H}}_1$ is made simply to provide an immediate intuition of the protocol through a significant example.} The dynamics encompassed by $\hat{{\cal H}}_1$ will be analyzed via the IF~\cite{informationflux}. In particular, within the framework of explicitly limited accessability stated above, we focus our attention on the evolution of qubit $1$. We thus move to the Heisenberg picture and consider the dynamics of the Pauli operators of this spin under the action of $\hat{{\cal H}}_1$. From now on, time-evolved operators will be indicated as $\hat{\cal O}(t)=\hat{\cal U}^\dag\hat{O}\,\hat{\cal U}$ with $\hat{\cal U}(t)=e^{-({i}/{\hbar})\hat{\cal H}_1t}$. A straigthforward calculation, based on the use of the operator expansion theorem, leads to the following decompositions of $\hat{\cal X}(t)$ and $\hat{\cal Y}(t)$ over sets of  $N$-spin operators~\cite{informationfluxijqi}
\begin{equation}
\label{evol1}
\begin{split}
\hat{\cal X}_1(t)&\!=\alpha_1(t)\hat{X}_1\!+\hat{Z}_1[\alpha_2(t)\hat{Y}_2\!+\cdot\cdot+\alpha_N(t)\hat{Z}_2\cdot\cdot\hat{R}_N],\\
\hat{\cal Y}_1(t)&\!=\beta_1(t)\hat{Y}_1\!+\hat{Z}_1[\beta_2(t)\hat{X}_2\!+\cdot\cdot+\beta_N(t)\hat{Z}_2\cdot\cdot\hat{S}_N],
\end{split}
\end{equation}
with $\hat{R}_N=\hat{X}_N$ and $\hat{S}_N=\hat{Y}_N$ ($\hat{R}_N=\hat{Y}_N$ and $\hat{S}_N=\hat{X}_N$) for odd (even) $N$. The time-dependent parameters $\alpha_i(t)$'s and $\beta_i(t)$'s ($i=1,...,N$) are functions of the coupling strength set $\{J_i\}$. For instance, we have $\alpha_1(t)\!=\!\sum_{l=0}^{\infty}[(2t)^l/l!]\delta_1^{(l)}$ with the recurrence formula $\delta_j^{(l)}\!=\!(-1)^{j}[J_{j-1}\delta_{j-1}^{(l-1)}+J_{j}\delta_{j+1}^{(l-1)}]$, $J_0\!=\!J_N\!=\!0$ and the initial conditions $\delta_j^{(0)}=0$ ($1$) for $j\ne 1$ ($j=1$). Therefore, $\alpha_1(t)$ depends on the full set $\{J_i\}$.

\begin{figure}[t]
{\bf (a)}\hskip4cm{\bf (b)}
\centerline{\psfig{figure=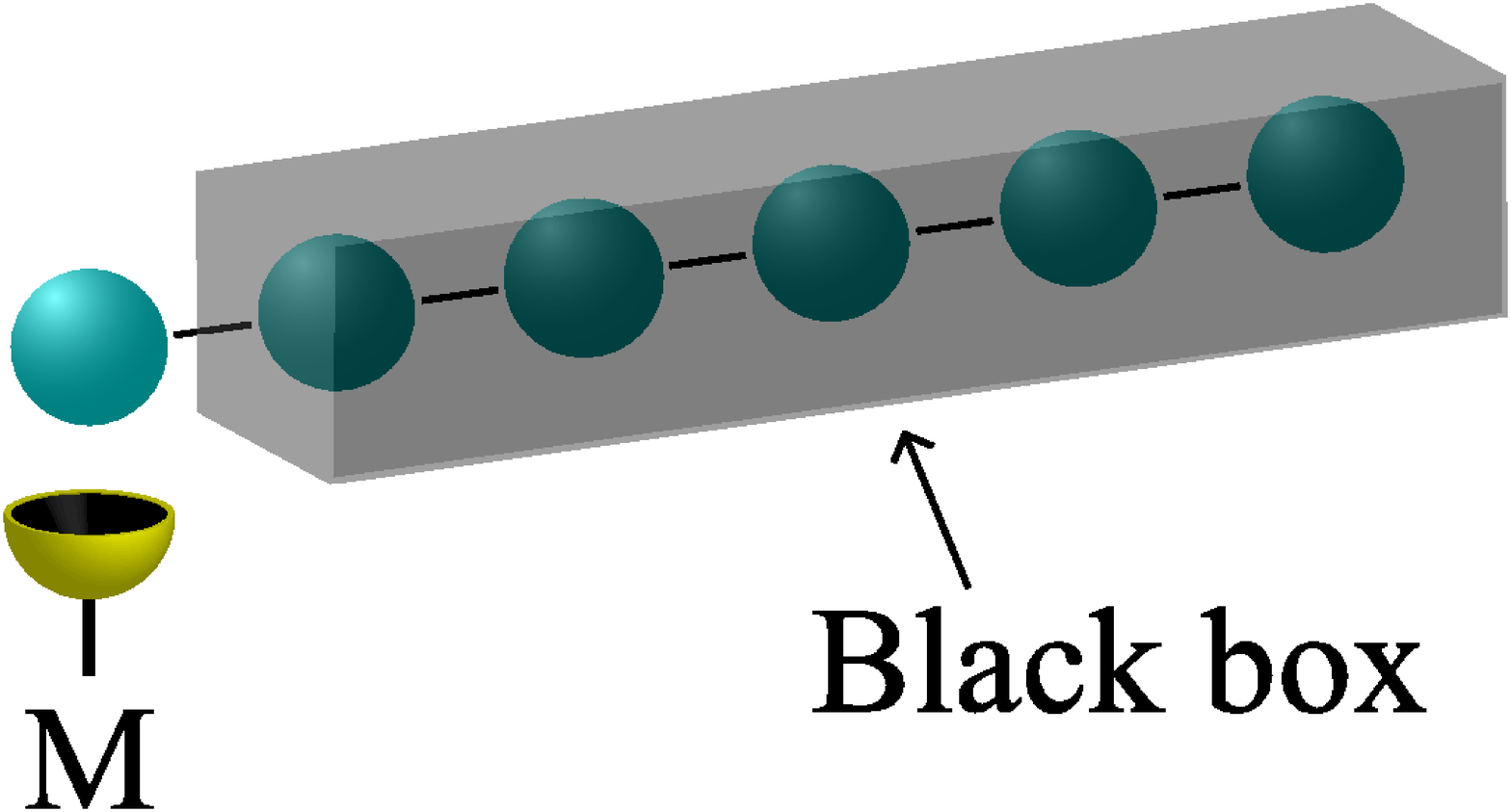,height=2cm}\hskip0.75cm\psfig{figure=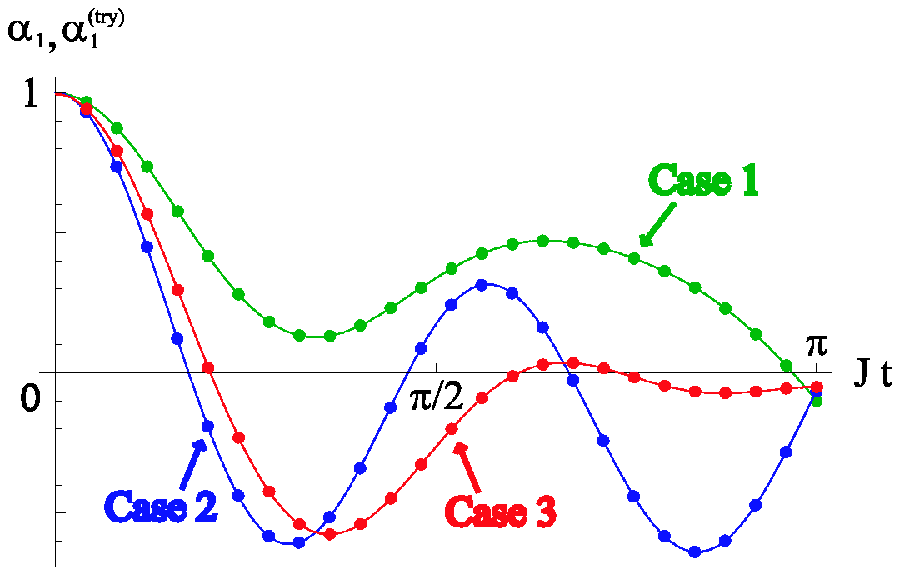,height=2.625cm}}
\caption{{\bf (a)} Sketch of the scheme for Hamiltonian tomography without state initialization, where $M$ is the measurement performed on spin $1$ and other spins are not accessible; {\bf (b)} Simulated dynamics of $\minivalmed{\hat{X}_1(t)}$, sampled at steps of $Jt=\pi/25$, under the action of $\hat{\cal H}_{1}$ with $N=6$ and $\{J_i/J\}=\{1.02,1.26,0.94,1.36,0.72\}$ (case 1), $\{J_i/J\}=\{1.49,0.80,1.02,0.69,1.28\}$ (case 2), and $\{J_i/J\}=\{1.30,0.80,1.23,0.75,0.96\}$ (case 3). The corresponding fits are performed using the trial function in Eq.~(\ref{try}), which is in excellent agreement with the behavior of the data.}
\label{plot1}
\end{figure}

Focusing, for the sake of argument, on the first of Eqs.~(\ref{evol1}), if we initialize spin $1$ in an eigenstate of $\hat{X}_1$, $\ket{\pm_x}_{1}=(\ket{0}\pm\ket{1})_{1}/\sqrt{2}$, we get $\minivalmed{\hat{X}_1(t)}\!=\!\pm\alpha_1(t)$. Analogously, if the initial state of spin $1$ is $\ket{\pm_y}_1\!=\!(\ket{0}\pm i\ket{1})_{1}/\sqrt{2}$, we would obtain $\minivalmed{\hat{Y}_1(t)}=\pm\beta_1(t)$. It is easy to see that, for the case of $\hat{{\cal H}}_1$, the recurrence formulas that determine $\alpha_{i}(t)$'s and $\beta_i(t)$'s are exactly the same, so that $\alpha_1(t)=\beta_1(t)$. This is due to the symmetric role played by the $\hat{X}_i\hat{X}_{i+1}$ and $\hat{Y}_i\hat{Y}_{i+1}$ terms in Eq.~(\ref{modelloXX}). As stated above, $\alpha_1(t)$ depends on the full set $\{J_i\}$. Therefore, in order to obtain information about all the coupling strengths within $\hat{\cal H}_1$, we need to determine the functional behavior of the expectation value of a single one-qubit operator, such as $\minivalmed{\hat{X}_1(t)}$, and make explicit the relation connecting it to $J_i$'s. As we show later, this can be done via  a simple post-processing step of the measured data.

At first sight, it might seem that spin $1$ should be properly prepared. However, this is clearly not the case. In fact, we just need to sample $\minivalmed{\hat{X}_{1}(t)}$ at successive instants of time, so that each $X$-projection we perform on spin 1 prepares it into the desired class of states. From that point on, $\minivalmed{\hat{X}_1(t)}\!=\!\pm\alpha_1(t)$ holds rigorously. The necessity of iterated state initializations is thus bypassed. A sketch of the scheme is presented in Fig.~\ref{plot1}{\bf (a)}. A second important observation is that, by having decoupled the evolution of $\minivalmed{\hat{\sigma}_1(t)}$ ($\sigma=X,Y$) from the explicit influences of the expectation value of operators involving spins from $2$ to $N$, {\it the initial state of the rest of the system might be completely arbitrary and unknown}. To the best of our knowledge, this feature is unique to our method. As no information is required on the dynamical aspects of the rest of the chain, our Hamiltonian tomography is performed with only minimal invasiveness on the many-body system.

In order to show the efficiency of the method and clarify its working principles, it is worth addressing a few explicit examples. We have generated random sets of coupling parameters (for a chain of $N=6$ spins) taken from a uniform distribution in the range $[0.5J,1.5J]$, where $J$ is an arbitrary constant. We have then simulated $X$-measurements on spin $1$ in a way so as to get a $25$-point sample of $\minivalmed{\hat{X}_1(t)}$ for each case, taking $t\in[0,\pi/J]$~\cite{timereasons} at steps of $\pi/(25J)$. The elements of three of such samples are shown in Fig.~\ref{plot1}{\bf (b)}. As a post-processing stage of our analysis, we now need to fit the points within each sample with a proper functional form which we take as a linear combination of trigonometric cosine functions of unknown amplitudes and frequencies. The choice of such basis of functions is not arbitrary and is somehow induced by the interference nature of the mechanism behind information-propagation across a spin-chain, as discussed in~\cite{bruder}. Moreover, this form can also be inferred from the functional form of $\alpha_1(t)$ in the particular case of sets $\{J_i\}$ allowing perfect state transfer~\cite{pst}. For the case at hand, we find that the trial function
\begin{equation}
\label{try}
\alpha_1^{\rm{(try)}}(t)={\cal A}\cos(\omega_{\cal A}t)+{\cal B}\cos(\omega_{\cal B}t)+{\cal C}\cos(\omega_{\cal C}t)
\end{equation}
is in excellent agreement with the behavior of the simulated data, as shown in Fig.~\ref{plot1}. By equating the amplitudes ${\cal A,B,C}$ and frequencies $\omega_{\cal A,B,C}$ to the functions of $J_i$'s entering into $\alpha_1(t)$, we have estimated $J_{i}/J$ ($i=1,..,5$) to be within $0.1\%$ of the values listed in the caption of Fig.~\ref{plot1}{\bf (b)} for each of the cases shown. {It is remarkable that, differently from previous proposals~\cite{burgarth}, the energy spectrum of the coupling Hamiltonian is required at no stage of the protocol.} For an ideal unitary evolution, less points within each sample are actually sufficient to estimate the parameters. For instance, we have obtained $J_i$'s with a good precision by considering only $10$ points and light computational effort for the fit. However, the plots presented here include $25$ points, as this helps in optimizing their visualization. Although the analysis above provides clear evidence that the accuracy of the protocol is almost insensitive to the particular choice of $J_i$'s, we have explicitly checked this feature by simulating the performance of the tomography protocol for several randomly generated sets, evaluating the average error associated with the retrieval of the corresponding coupling parameters. We have studied chains of up to $N=8$ spins and found an average relative error always smaller than $0.3\%$~\cite{comptime}. 

As previously mentioned, the method can be extended to a more general Hamiltonian model. In fact, let us consider the following interaction model, which {\it does not preserve the total number of excitations in the system}
\begin{equation}
\hat{{\cal H}}_2=\sum^{N-1}_{i=1}(J_{X,i}\hat{X}_{i}\hat{X}_{i+1}+J_{Y,i}\hat{Y}_{i}\hat{Y}_{i+1}).
\end{equation}
The evolution of $\hat{X}_1$ and $\hat{Y}_1$ under $\hat{\cal H}_2$ can still be written as in Eqs.~(\ref{evol1}). The coefficients $\alpha_k$'s and $\beta_k$'s depend, in this case, on two disjoint and alternate sets of parameters $J_{\sigma,i}$'s (this result has been recently exploited in Ref.~\cite{matryoshka}). For instance, $\alpha_k$'s ($\beta_k$'s) depend only on $J_{X,k}$'s with even (odd) $k$ and $J_{Y,k}$'s with odd (even) $k$. Actually, the same recurrence formulas used above also hold in the present case. We can thus perform the tomographic protocol twice: first we consider $\ket{\pm_x}_1$ as the initial state of the first qubit and evaluate $\minivalmed{\hat{X}_1(t)}$. Then, we estimate  $\minivalmed{\hat{Y}_1(t)}$ from the initial state $\ket{\pm_y}_1$. In this way we obtain information on both sets of parameters and we can reconstruct the complete set $\{J_{\sigma,i}\}$. It is important to stress that this excitation-non-preserving case cannot be analyzed by means of protocols such as the ones in Ref.~\cite{burgarth}, which critically rely on the condition of excitation-conservation.

\begin{figure}[t]
{\bf (a)}\hskip4cm{\bf (b)}
\centerline{\psfig{figure=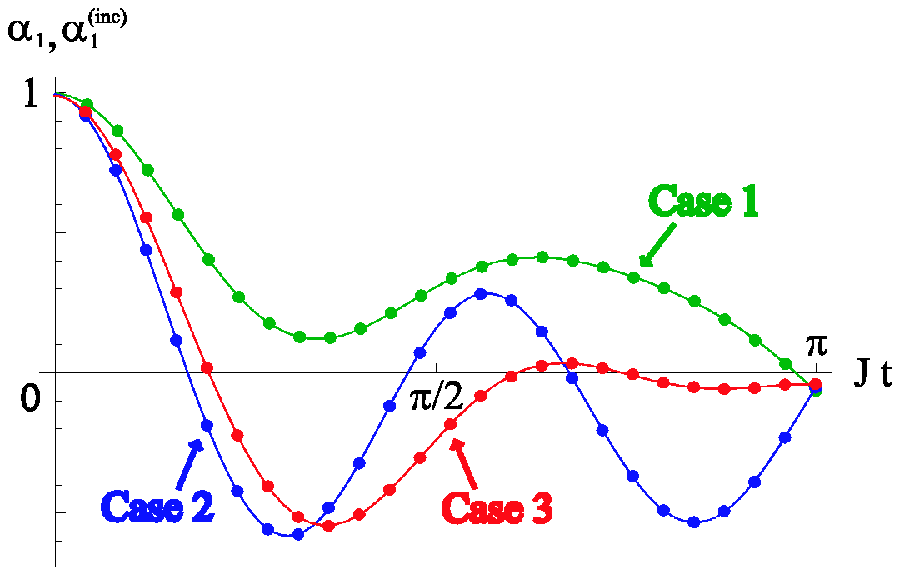,height=2.625cm}\hskip0.25cm\psfig{figure=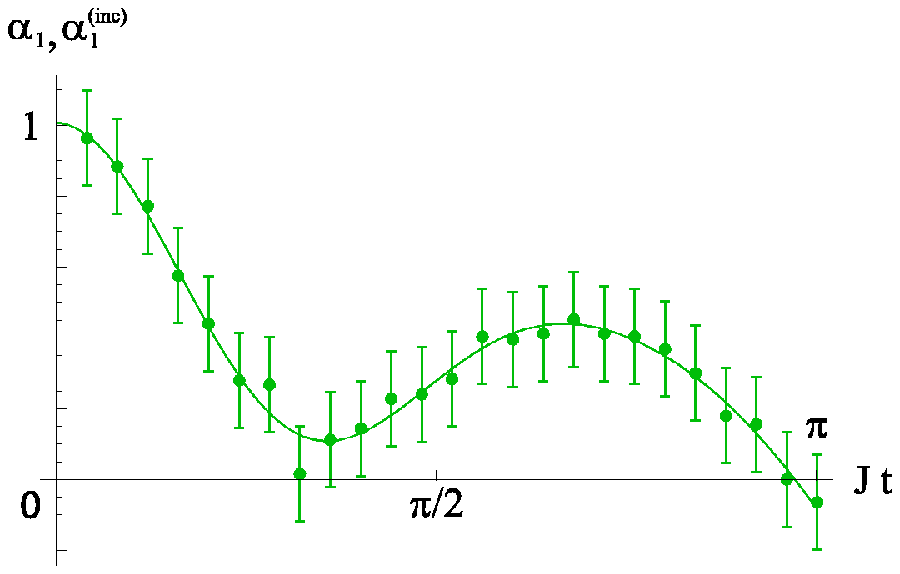,height=2.625cm}}
\caption{{\bf (a)} Simulated dynamics of $\minivalmed{\hat{X}_1(t)}$ in the presence of incoherent coupling of the system with an environment. We have considered $N=6$, $\gamma/J=0.5$, $\Gamma/J=0.2$, $\nbar=0.01$, and the sets $\{J_i/J\}$ listed in the caption of Fig.~\ref{plot1}{\bf (b)}. The corresponding fits are performed using the trial function in Eq.~(\ref{try2}), which is in excellent agreement with the behavior of the data; {\bf (b)} Simulated dynamics of $\minivalmed{\hat{X}_1(t)}$ in the same setting, with a finite number of measurements $n_{meas}=500$ and the set $\{J_i\}$ corresponding to case 1 of Fig.~\ref{plot1}{\bf (b)}.} 
\label{plot2}
\end{figure}

The tomographic scheme we propose is able not only to provide information on the unitary dynamics of the elements of the chain but also to estimate the effects of incoherent coupling of the system with an environment. In what follows, we describe how the influence of dissipation and dephasing on a spin-chain can be retrieved from our formal apparatus. {For the sake of argument, we consider again model $\hat{{\cal H}}_1$ although, we remark, the results will be valid for the more general Hamiltonian $\hat{{\cal H}}_2$ as well.} We assume weak-coupling conditions between each spin of the chain and its own bath, modeled as an ensemble of bosonic modes. We consider the effects of both amplitude and phase damping channels over the chain. Using the operator-sum representation, the evolution of an initial state $\rho_c$ of the whole chain under the effect of one of such channels is given by $\varrho_c(\tau)=\sum_{\mu}\hat{K}^{\mu}(\tau)\rho_c\hat{K}^{\mu\dag}(\tau)$. Here, $\{\hat{K}^\mu(\tau)\}$ is the set of time-dependent Kraus operators such that $\sum_{\mu}\hat{K}^{\mu\dag}(\tau)\hat{K}^\mu(\tau)=\openone$ and $\tau$ is the interval during which the channel is acting~\cite{nielsenchuang}. The formal description of a single-spin amplitude damping process in a bath in equilibrium at a finite temperature is described by the set $\{\hat{K}^\mu_{i}(\tau)\}=\{\hat{A}^0_i,\hat{A}^1_i,\hat{A}^2_i,\hat{A}^3_i\}$, where $\hat{A}^0_{i}=\sqrt{p}(\miniproj{0}{0}+e^{-\Gamma\tau/2}\miniproj{1}{1})$, $\hat{A}^1_{i}=\sqrt{p(1-e^{-\Gamma\tau})}\,\miniproj{0}{1}$, $\hat{A}^2_{i}=\sqrt{1-p}(e^{-\Gamma\tau/2}\miniproj{0}{0}+\miniproj{1}{1})$, $\hat{A}^3_{i}=\sqrt{(1-p)(1-e^{-\Gamma\tau})}\,\miniproj{1}{0}$ with $p=(\nbar+1)/(2\nbar+1)$ and $\nbar$ the average phonon number of the bath, assumed to be the same for each spin. For a dephasing channel, on the other hand, we have $\{\hat{K}^\mu_{i}(\tau)\}=\{\hat{D}^0_i,\hat{D}^1_i\}$ with $\hat{D}^0_i=\sqrt{(1+e^{-\gamma\tau})/2}\,\hat{\openone}_i$, $\hat{D}^1_i=\sqrt{(1-e^{-\gamma\tau})/2}\,\hat{Z}_i$. $\Gamma$ and $\gamma$ are the rates of amplitude and phase damping, respectively. {Our approach is to intersperse the unitary evolution $\varrho_c(t)=\hat{\cal U}\rho_c\,\hat{\cal U}^\dag$ (induced by $\hat{{\cal H}}_1$) and the non-unitary dynamics (resulting from the integration of the Lindblad equation corresponding to a given noise channel), each lasting for small time interval $\Delta{t}$}. We randomly select the spin upon which apply the operator-set $\{\hat{A}^\mu_i\}$ or $\{\hat{D}^\mu_i\}$. The results of the simulated measurements are then averaged over the collection of noise-occurrence patterns (runs), in order to guarantee the faithful unraveling of the open quantum dynamics. Finally. Hamiltonian tomography is performed. An elegant and effective description of open dynamics can be given in terms of IF formalism considering that the action of a set of Kraus operators on $\hat{\sigma}_j$ is obviously given by $\hat{\cal O}(\tau)\!=\!\sum_{\mu}\hat{K}^{\mu\dag}(\tau)\hat{O}\hat{K}^{\mu}(\tau)$. Therefore, for a dephasing channel acting on qubit $i$, we have
\begin{equation}
\hat{\cal X}_i=e^{-\gamma \tau}\hat{X}_i\, ,  \;\;\;\;\hat{\cal Y}_i=e^{-\gamma \tau}\hat{Y}_i\, , \;\;\;\;\hat{\cal Z}_i=\hat{Z}_i.
\end{equation}
It is immediate to recognize that this results in the change $\alpha_i(t)\rightarrow{e}^{-\gamma \tau}\alpha_i(t)$ [$\beta_i(t)\rightarrow{e}^{-\gamma{\tau}}\beta_i(t)$] in the decomposition of $\hat{\cal X}_1(t)$ [$\hat{\cal Y}_1(t)$] in Eqs.~(\ref{evol1}), with all the other terms unmodified. The description in Eqs.~(\ref{evol1}) will thus remain formally the same, together with all the qualitative results presented in the unitary case. Computationally, the analysis performed with this method is faster than the one based on density matrix evolution (the computational time grows as $N^2$). For an amplitude damping channel, we have
\begin{equation}
\begin{split}
&\hat{\cal X}_i=e^{-\frac{\Gamma \tau}{2}}\hat{X}_i\, , \;\;\;\hat{\cal Y}_i=e^{-\frac{\Gamma \tau}{2}}\hat{Y}_i,\\
&\hat{\cal Z}_i=(1-e^{-\Gamma \tau})(2p-1)\hat{\openone}_i+e^{-\Gamma \tau}\hat{Z}_i.
\end{split}
\end{equation}
The decomposition of $\hat{\cal X}_1(t)$ [$\hat{\cal Y}_1(t)$] after the action of the channel is therefore no more restricted to the operator set used in Eqs.~(\ref{evol1}) but involves a larger one. For our numerical study, we have taken $\gamma/J=0.5$, $\Gamma/J=0.2$ and $\nbar=0.01$. In Fig.~\ref{plot2}{\bf (a)} we present the simulated points obtained for a chain of $6$ qubits with the sets $\{J_i\}$ previously considered and $100$ runs.
We have found that the trial function
\begin{equation}
\label{try2}
\alpha_1^{\rm{(inc)}}(t)\!=\![{\cal A}\cos(\omega_{\cal A}t)+{\cal B}\cos(\omega_{\cal B}t)+{\cal C}\cos(\omega_{\cal C}t)]e^{-\tilde{\gamma}t}
\end{equation}
with $\tilde{\gamma}$ an effective rate depending on $\gamma$ and $\Gamma$, is in excellent agreement with the behavior of the simulated data, as shown in Fig.~\ref{plot2}{\bf (a)}. Surprisingly, the amplitudes ${\cal A,B,C}$ and frequencies $\omega_{\cal A,B,C}$ are the same as in the ideal case. The only net effect of noise in $\minivalmed{\hat{X}_1(t)}$ is the damping of the oscillations. Also in the presence of incoherent coupling of the system with an environment, our Hamiltonian tomography protocol works well. We have proved it by obtaining ${\cal A,B,C}$ and $\omega_{\cal A,B,C}$ from the fits in Fig.~\ref{plot2}{\bf (a)} and estimating $J_{i}/J$. The results are within $4\%$ of the original values, for each of the cases analyzed. 

In order to predict the performance of our tomography process in a way so as to be closer to realistic conditions, we have considered the error due to the finite number of measurements $n_{\text{meas}}$ performed to evaluate, each time, the required expectation values. We have estimated the parameters $J_i$'s in the three cases above, including noise, for $n_{\text{meas}}=500$. The results are within $9\%$ of the expected values, for each of the cases analyzed. This error can be reduced by increasing the number of measurements per sampling time. A set of simulated outcomes of the measurements is presented in Fig.~\ref{plot2}{\bf (b)}.

Finally, for the sake of completeness, we have analyzed the case with additional (unknown) spurious terms in the coupling model, such as local magnetic fields along the $z$-axis or interaction terms proportional to $\hat{Z}_i\hat{Z}_{i+1}$. This would get our study closer to a true experimental situation where unwanted ``engineering'' defects could affect a Hamiltonian. In the limit of small influences ( $\simeq0.1J$) {of} the additional terms on $\hat{{\cal H}}_1$ or $\hat{{\cal H}}_2$ and for only a finite number of measurements being performed, our protocol can estimate the parameters $J_i$'s with a $10\%$ error. This value is comparable to the one obtained without spurious couplings and an equally finite sampling, which shows that the effects of the additional terms is very small. 

We have proposed a scheme for the tomography of a wide class of interaction Hamiltonians. Our method is designed to work in a scenario of restricted accessability to the components of a spin chain. It allows the identification of coupling parameters through the temporal dynamics of a single spin. As no initial state preparation is necessary, measurements can be performed by interspersing the system's evolution. Besides data acquisition, only a simple post-processing step is necessary: no conservation law associated to the interaction or {\it a priori} knowledge on the state of the system is required. Even when a spin chain is affected by environmental influences, our Hamiltonian tomography remains possible and reliable. In order to widen the class of Hamiltonians that can be assessed with our scheme, including the case of external local fields applied to the system, we can speculate various strategies, such as dynamical changes of measurement basis. A general theory for Hamiltonian tomography will be an exciting extension of the work presented in this Letter. Given the crucial role that proper coupling patterns have in the interference effects behind quantum many-body phenomena, non-demanding diagnostic methods are important tools which need to be developed. Our proposal contributes to this task in a significant way. 

We thank T. J. G. Apollaro, D. Ballester, D. Burgarth, R. Filip, N. Lo Gullo and F. Plastina for discussions. We acknowledge support from the UK EPSRC and QIPIRC. M.P. is supported by EPSRC (EP/G004579/1).


\begin{thebibliography}{99}
\bibitem{antiferromagnetic} The condition of anti-ferromagnetic coupling can be dropped if the sign of all the parameters is known.
\bibitem{timereasons} The time interval was chosen so as to allow the determination of $\alpha_1(t)$ without worry about environmental effects. With our choice of parameters, these kick in at later times.
\bibitem{comptime} The computational time required to get each fitting was within a few minutes on a 2.4 GHz desktop machine with 2 Gb RAM.
\end{thebibliography}
\end{document}